\documentclass[]{spie}

\usepackage{amsmath,amsfonts,amssymb}
\usepackage{graphicx}
\usepackage{rotating}
\usepackage[colorlinks=true, allcolors=blue]{hyperref}
\usepackage[caption=false,font=footnotesize]{subfig}

\title{Performance of the Nonlinear Curvature Wavefront Sensor as a Function of Scintillation Strength}

\author[a,b]{Stanimir Letchev}
\author[a]{Justin R. Crepp}
\author[a]{Caleb G. Abbott}
\author[a]{Ryan Hersey}
\author[a]{Matthew Engstrom}
\author[a]{Nicholas Baggett}
\affil[a]{University of Notre Dame, Physics and Astronomy, Notre Dame, IN 46556, USA}
\affil[b]{Max-Planck Institute for Astronomy, Königstuhl 17, 69117 Heidelberg, Germany}

\authorinfo{Further author information: (Send correspondence to S.L.)\\S.L.: E-mail: letchev@mpia.de}

\begin{document} 
\maketitle

\begin{abstract}
Local amplitude aberrations caused by scintillation can impact the reconstruction process of a wavefront sensor (WFS) by inducing a spatially non-uniform intensity at the pupil plane. This effect is especially relevant for the commonly-used Shack-Hartmann WFS (SHWFS), which can lose slope information for portions of the beam where the signal is faint, leading to reduced reconstruction performance and eventually total failure as the level of scintillation increases. An alternative WFS is needed for such conditions. The nonlinear curvature wavefront sensor (nlCWFS) has been shown to achieve better sensitivity compared to the SHWFS under low light levels. Additionally, the nlCWFS has demonstrated the ability to maintain its sensitivity in the presence of scintillation, using amplitude aberrations to help inform the reconstruction process, rather than hinder. Experiments to date have thus far only shown reconstruction results for a single scintillation value. Building upon previous simulations and laboratory experiments, we have built a testbed to quantify the effects of varying scintillation strength on the wavefront reconstruction performance of the nlCWFS compared to an equivalent SHWFS. In this paper, we present results showing the difference in performance between the nlCWFS and SHWFS as a function of relative flux and scintillation strength.
\end{abstract}


\section{Introduction}\label{sec:intro}

The nonlinear curvature wavefront sensor (nlCWFS) is an intrinsically sensitive device that has been shown to maintain its performance while in the presence of moderate scintillation \cite{guyon_10,crass_thesis,mateen_15,Crepp_20}. Crepp et al. 2020 found nearly an order of magnitude improvement in the faintest flux levels that a nlCWFS can operate compared to an equivalent SHWFS. However, the Crepp et al.\ 2020 experiment was performed with only a single scintillation index. Using numerical simulations, Potier 2023 predicts that the relative performance of the nlCWFS over the SHWFS should continue to improve with increasing scintillation strength.\cite{Potier_thesis}

Building upon the Crepp et al.\ 2020 and Potier 2023 studies, these proceedings present preliminary results from a laboratory test-bed that was designed to experimentally quantify the effects of varying scintillation strength on the wavefront reconstruction accuracy of the nlCWFS compared to the SHWFS. Section \ref{sec:methods} describes the experimental methods and test-bed used for measurements. Section \ref{sec:results} presents results of wavefront error as a function of flux level and scintillation strength. Section~\ref{sec:summary} provides a summary and concluding remarks.
\section{Methods}\label{sec:methods}

\subsection{Experimental Design}\label{sec:experiment}

\begin{figure*}
    \centering
    \includegraphics[width=\textwidth]{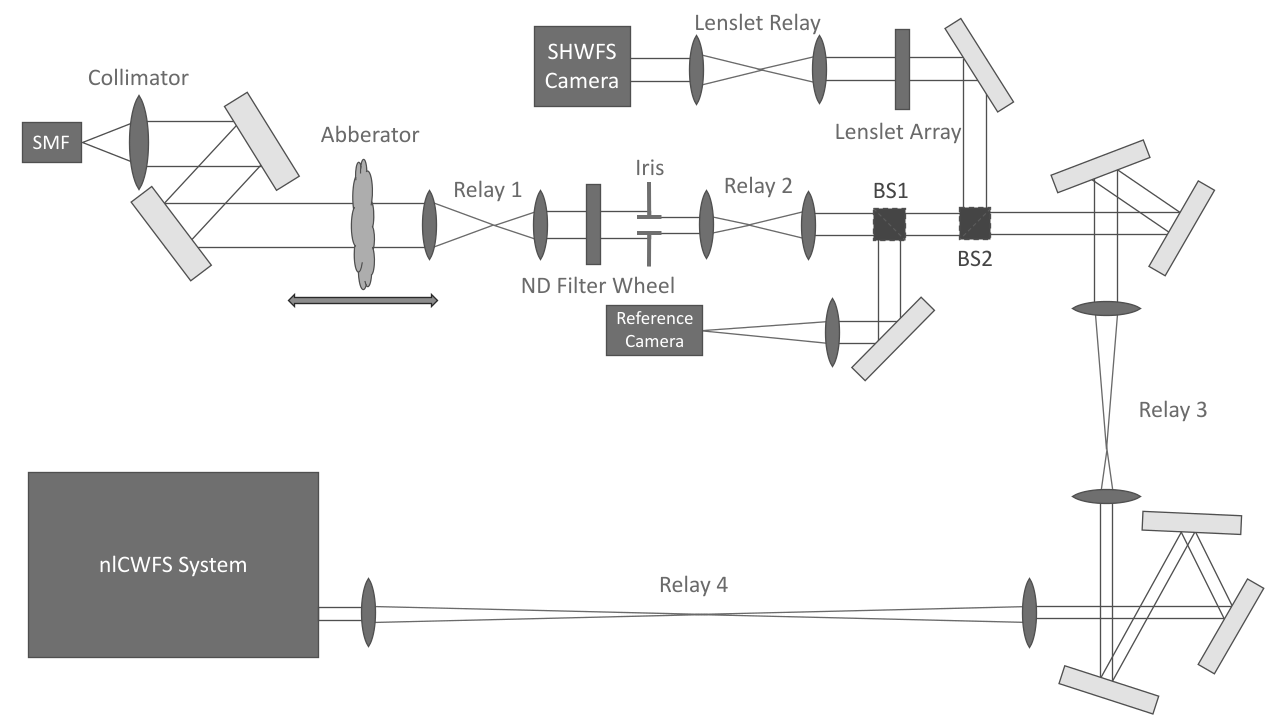}
    \caption{Optical layout of the scintillation test-bed. See text for discussion.}
    \label{fig:component_layout}
\end{figure*}

The experimental layout is based on that of Crepp et al.\ 2020 \cite{Crepp_20} with several upgrades introduced to reduce experimental uncertainty. Figure~\ref{fig:component_layout} shows a simplified diagram of the components layout. A variable-intensity monochromatic laser ($\lambda=532$ nm) illuminates the system via a single mode fiber (Thorlabs P1-460B-FC-1).  Aberrations are introduced in a collimated space using a single phase plate that had been sprayed with acrylic, which simulates a Kolmogorov atmospheric turbulence spectrum \cite{thomas_04}. To create a variety of wavefront aberrations, different regions of the plate are sampled by translating the plate in directions orthogonal to the beam, using a remotely-operated translation stage (Thorlabs PT1-Z8). The phase aberrations are then optically relayed using a two-lens system to an iris set to a diameter of 2 mm (Thorlabs SM1D12) that served as the ``telescope'' aperture. The $z$-distance of the plate was adjusted to vary the level of phase-induced amplitude aberrations (scintillation). The relay was designed to allow for the aberration plate to be placed directly conjugate to the pupil to introduce very low scintillation as a special case.

Following the iris, another relay is used to expand the beam to 4 mm and create a conjugate pupil plane for use by other system elements such as a deformable mirror. A series of two low-WFE beam-splitters divert light into three channels: two WFSs (SHWFS and nlCWFS) and a reference camera (Thorlabs CS2100M-USB) that is used for brightness calibration ($\S$\ref{sec:brightness_cal}). The two different WFSs are used to measure aberrations contemporaneously.

The SHWFS channel consists of a lenslet array (Thorlabs MLA150-7AR-M) placed at the conjugate pupil and a relay lens system. The relay was needed because the lenslets' focal length is too short given mechanical constraints in front of the detector. The nlCWFS channel consists of a series of relay lenses to reduce the beam size to 0.5 mm and create a conjugate pupil distance of 500 mm; this latter feature creates enough space for opto-mechanics while generating sufficient path-length diversity for wavefront reconstruction. The four nlCWFS measurement plane images were relayed onto the same detector using a series of beam splitting optics.

To compare sensors, both the nlCWFS and SHWFS each use a QHYCCD QHY294M Pro camera, thus allowing for common read noise, electronics, pixel pitch, and other characteristics. Although the two camera sensors are not perfectly identical, equal exposure time, binning, and gain settings were used to minimize systematic effects. To help maintain equivalence between the two WFSs, a single cut-out of $1000 \times 1000$ pixels was used for the SHWFS, and four cut-outs of $500 \times 500$ pixels each were used for the nlCWFS measurement planes. This sampling is equivalent to $\approx$ 100 pixels across the nlCWFS pupil plane and 26 lenslets with $\approx$ 8 pixels across each lenslet for the SHWFS pupil.

\subsection{Brightness Variation and Calibration}\label{sec:brightness_cal}

Since the source intensity could only be varied by approximately one order of magnitude using the laser current, a series of neutral density (ND) filters were incorporated into the beam to increase overall dynamic range of the experiment. The ND filters were installed in an electronic filter wheel, which allowed for remote control through an automated data collection program ($\S$\ref{sec:data}). The filters used had values of ND2, ND3, ND4, and ND5, corresponding to reductions in brightness by factors of $10^2, 10^3, 10^4,$ and $10^5$ respectively.

Relative flux in each WFS channel was calibrated by measuring photo-electrons, taking into camera gain and exposure time. As an extra precaution and means for checking calibration, an additional reference camera was incorporated into the experiment for flux measurements. Background contamination was minimized during data acquisition using baffling and environmental control and removed to negligible levels in post-processing.

\begin{figure}
     \centering
     \subfloat[\label{fig:low_scin_pupil}]{
         \centering
         \includegraphics[width=0.45\textwidth]{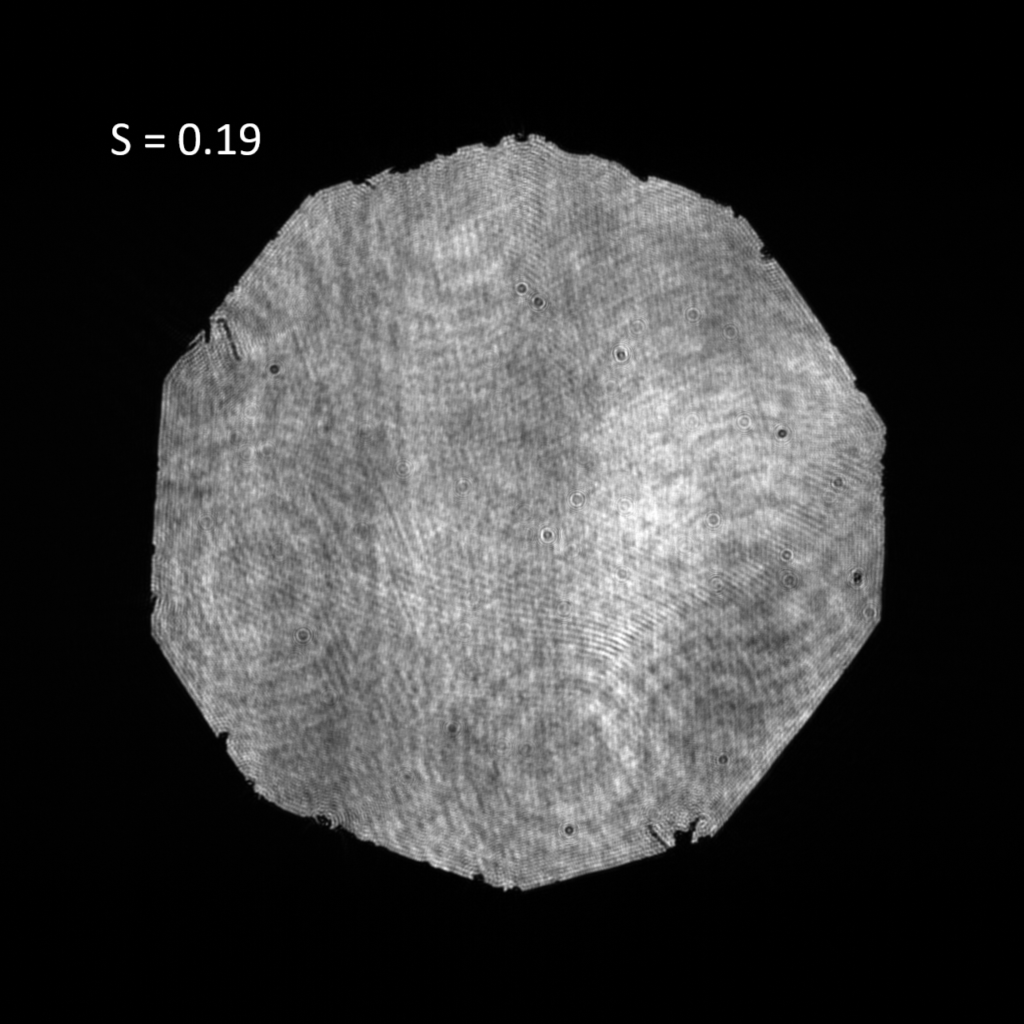}
     }
     \hfill
     \subfloat[\label{fig:high_scin_pupil}]{
         \centering
         \includegraphics[width=0.45\textwidth]{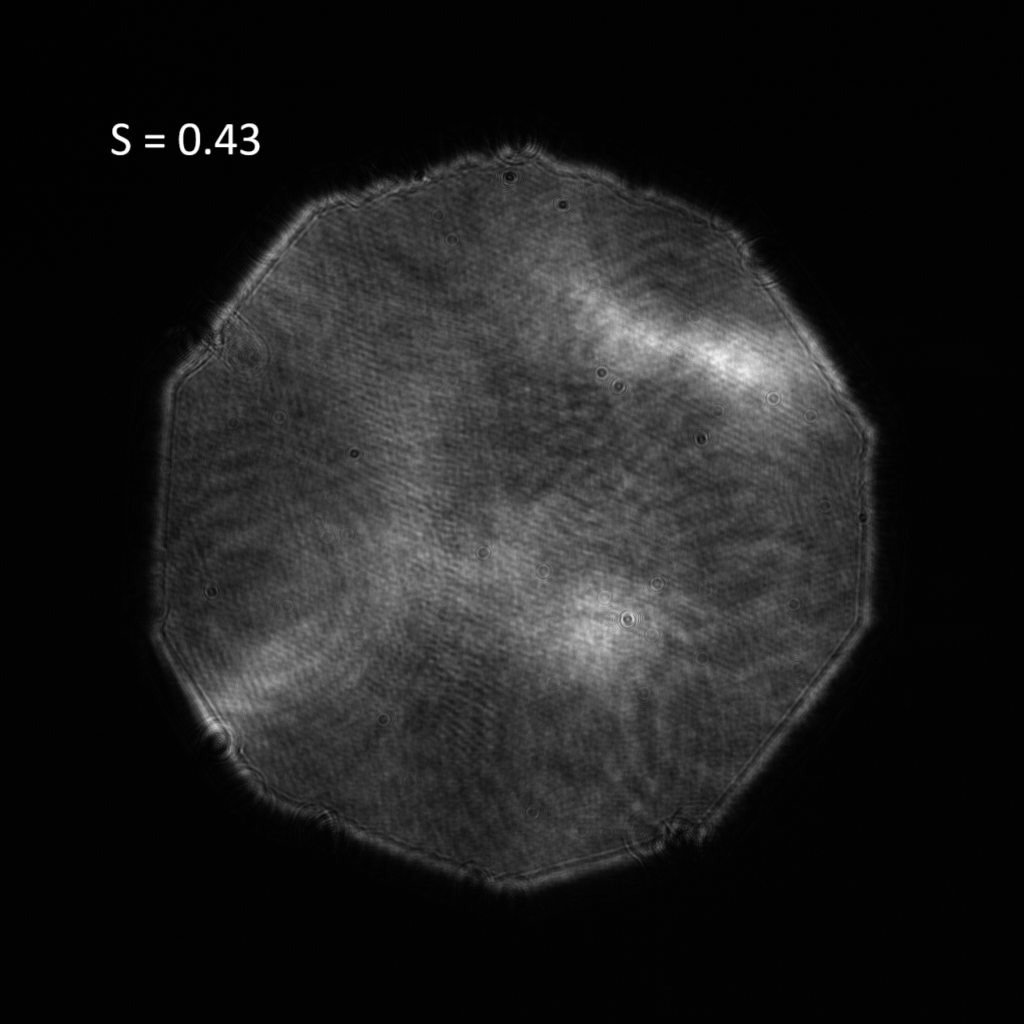}
     }
        \caption{Intensity of a conjugate pupil when placing the aberration plate (a) close to and (b) far from the pupil plane. The left image exhibits a low amount of induced scintillation ($S = 0.19$), because the phase aberrations have not induced significant amplitude variations. The right image exhibits a larger amount of induced scintillation ($S = 0.43$), because the phase aberrations have propagated further along the optical axis causing more internal beam interference. In both cases, higher-frequency features are due to aberrations introduced by the relay and beam-splitting optics. The dodecahedral shape of the pupil is due to the optical iris used to define the synthetic telescope aperture.}
\end{figure}

\subsection{Data Collection and Analysis}\label{sec:data}

To study the relative performance of the nlCWFS and SHWFS as a function of scintillation strength, a series of data collection runs were conducted at different scintillation values. The level of scintillation experienced in each experiment is quantified by the scintillation index ($S$), 
\begin{equation}
    S=\frac{\sqrt{\langle I^2 \rangle - \langle I \rangle^2}}{ \langle I \rangle}
\end{equation}
where $I=I(x,y)$ represents the measured intensity at the system pupil and brackets $<>$ indicate the spatial average.\footnote{Note that the equation from Crepp et al. 2020 calculates the value for $S^2$.} A value of $S = 0$ is indicative of no scintillation, whereas values approaching or exceeding $S = 1$ indicate strong scintillation. Different locations on the phase plate were studied for each $S$ value ($z$-distance) to build statics that average over randomly selected aberration maps.

As the scintillation index was varied, the performance of each sensor was tested as a function of incident flux. The intensity of the laser source illuminating the experiment was incrementally reduced until both sensors were clearly past the point of failure. To compare the experimental results to Crepp et al.\ 2020, a plot of residual wavefront error (WFE) versus relative flux was generated for each scintillation index. The plots are shown on a logarithmic scale in brightness to help identify any breaks in the photon-noise power law. 

For efficiency purposes, a custom LabVIEW program was developed that communicated with each camera, the laser controller, filter wheel, and translation stage that moved the aberration plate. The LabVIEW program automatically recorded images from each camera at each wavefront location, the exposure time, laser current, and filter wheel orientation. A typical data run comprised 5 - 10 different aberration maps, 45 brightness values (combination of laser current and ND filters), and 10 repeated exposures in each state for statistical robustness.

\subsection{Wavefront Reconstruction}\label{sec:reconstruct}

The SHWFS wavefront reconstruction used the same algorithm \cite{Antonello_14} as in the Crepp et al.\ 2020 study \cite{Crepp_20}, which was also used in \cite{Lechner_19} to study scintillation. The algorithm relies on a calibration image, which defines centroid offsets taken with the aberration disk removed from the beam prior to each data run. For the nlCWFS, all four measurement plane images were placed onto the same camera array. Measurement plane locations were placed symmetrically about the optical system pupil with distances selected using the approach detailed in Letchev et al.\ 2023\cite{letchev_22}. These initial locations were used as a starting point for the experiment and were further adjusted to optimize performance. Ultimately, distances of $Z_{\rm near}= \pm 10$ mm and $Z_{\rm far}= \pm 50$ mm were used. 

The nlCWFS also used the same reconstruction method used in Letchev et al.\ 2023, with small modifications made to accommodate experimental data. The reconstruction algorithm also used the same \textit{lspv} phase-unwrapping program from the commercially-available WaveProp and AOTools packages \cite{Brennan:16,Brennan:17}. The nlCWFS was calibrated by identifying centroids for each of the four measurement planes using a flat incoming wavefront (no aberration disk) prior to each data collection run. Centroid locations were calculated using a simple intensity weighted average method. See Abbott et al. 2024 in this SPIE proceedings for more advanced methods. 

\begin{figure}
     \centering
     \subfloat[\label{fig:low_scin_rms}]{
         \centering
         \includegraphics[width=\textwidth]{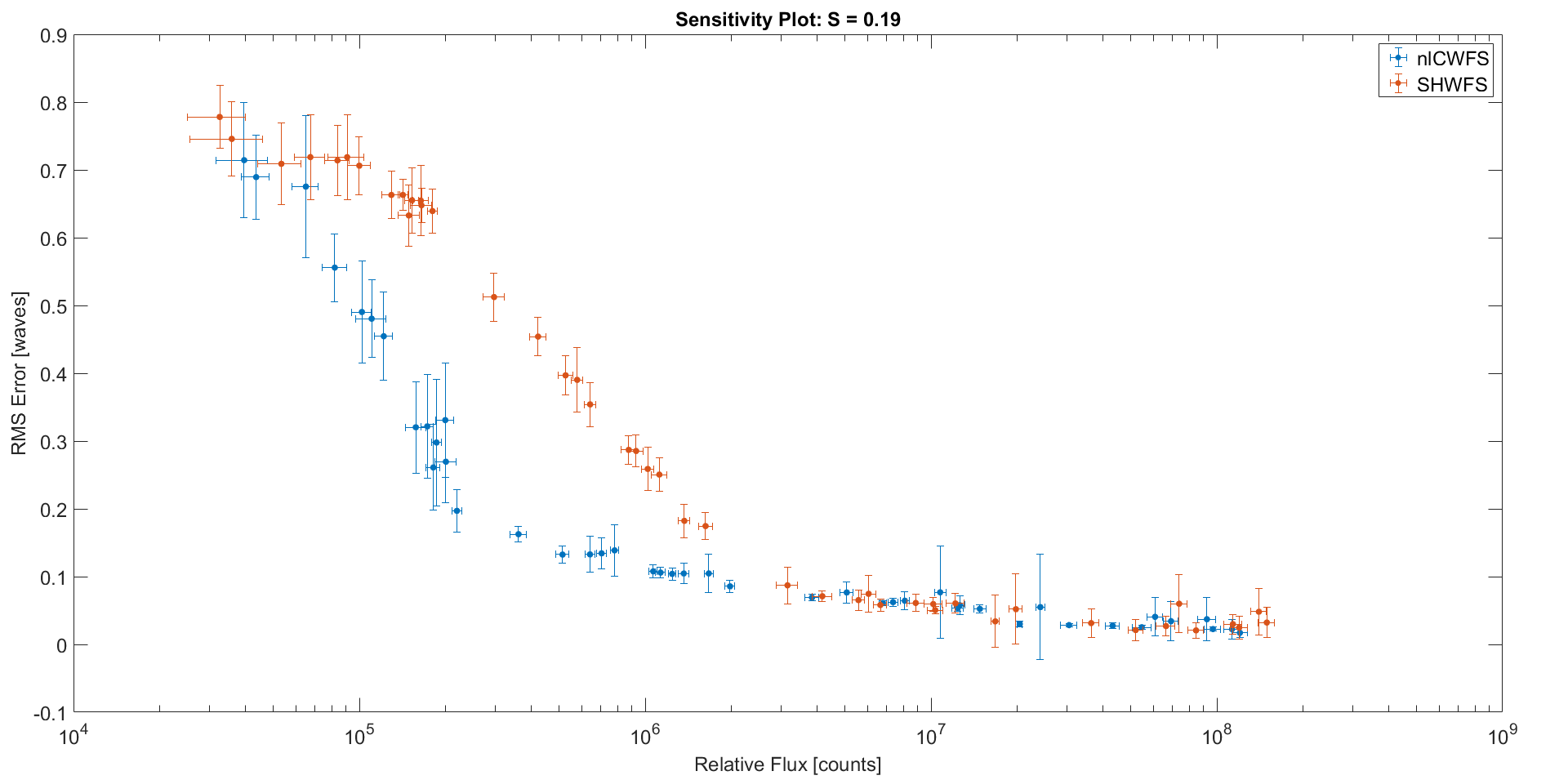}
     }
     \hfill
     \subfloat[\label{fig:high_scin_rms}]{
         \centering
         \includegraphics[width=\textwidth]{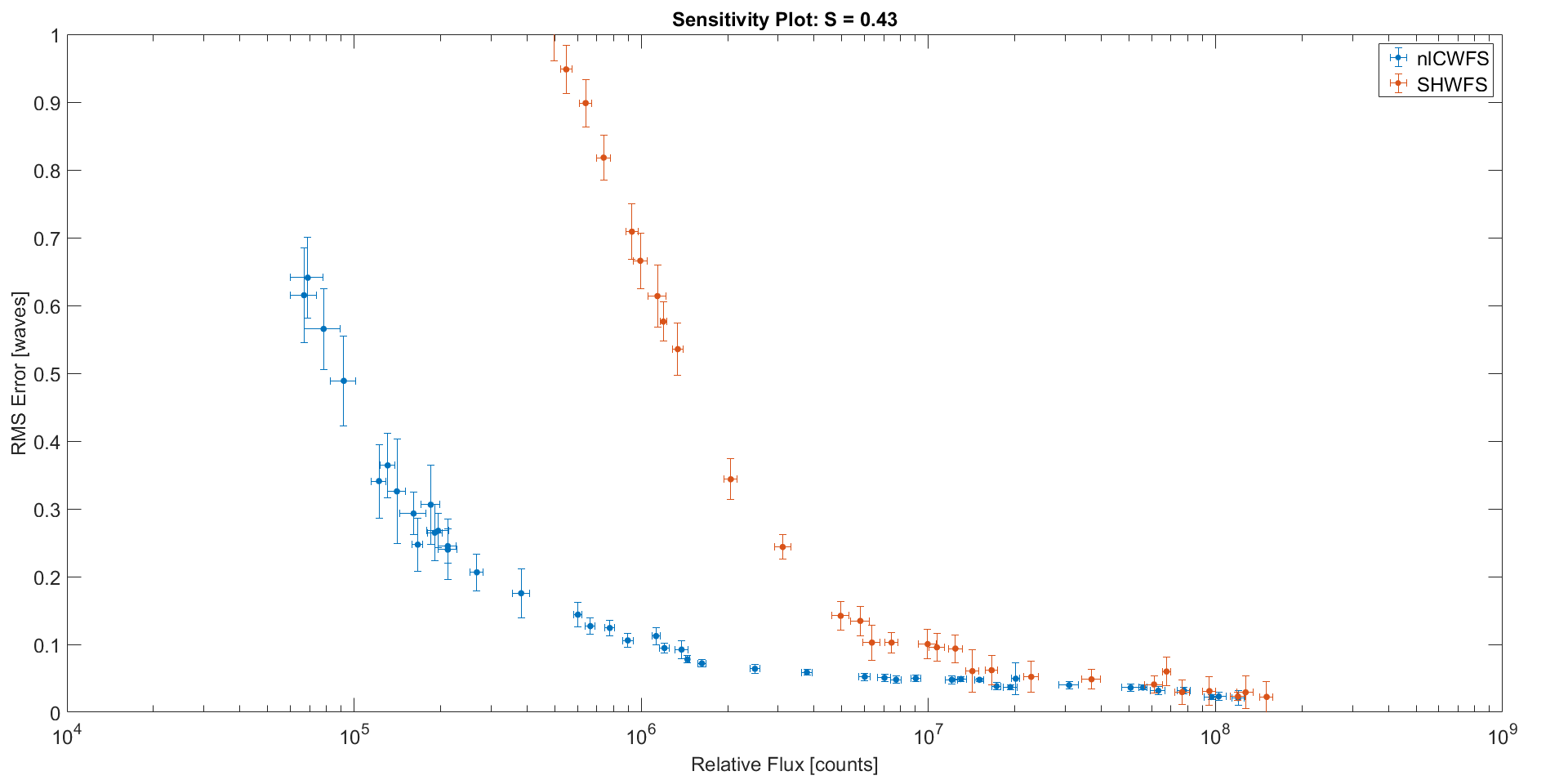}
     }
     \caption{Results of decreasing laser brightness using the nlCWFS (blue) and SHWFS (orange) with (a) low and (b) high levels of scintillation. In both cases, the SHWFS shows an earlier turn-off point in relative flux compared to the nlCWFS. In addition to showing that the nlCWFS is more sensitive than the SHWFS, this experiment demonstrates that the nlCWFS is able to maintain its performance across a relatively broad range of scintillation values.}
     \label{fig:extremes_scin_rms}
\end{figure}
\section{Experimental Results}\label{sec:results}

To verify the data collection and analysis techniques, as well as provide a baseline with which to compare other results, the first data run involved placing the phase aberration plate almost exactly at a conjugate pupil ($z \approx 0$). At this location, the scintillation index was estimated to be $S = 0.19$. A representative pupil image is shown in Figure \ref{fig:low_scin_pupil}. Intensity variations are minimal. 

When the aberration plate is moved along the $z$-axis, phase-to-amplitude conversion creates larger intensity variations, causing the scintillation index to increase (Fig.~\ref{fig:high_scin_pupil}). With this second more challenging data set, we estimate $S = 0.43$, meaning that variations in intensity are $\approx 43$\% of the average intensity.  

Figure~\ref{fig:low_scin_rms} and Figure~\ref{fig:high_scin_rms} show reconstruction results for each data set, plotting RMS WFE versus relative flux. As the laser brightness decreases, the RMS WFE increases until the point that each sensor eventually fails to reliably reconstruct the wavefront. We find that the nlCWFS is intrinsically more sensitive than a comparable SHWFS. This result is demonstrated by the fact that the nlCWFS outperforms the SHWFS even then the scintillation index is vanishingly small. 

We find that the nlCWFS is also less susceptible to increasing scintillation levels compared to the SHWFS, corroborating the results from Crepp et al. 2020.\cite{Crepp_20} In the case of stronger scintillation, the SHWFS curve moves to the right, requiring more light to achieve the same accuracy. Meanwhile, the nlCWFS curves do not change much between $S = 0.19$ and $S = 0.43$. For example, at a relative flux of $2\times10^{5}$, the RMS WFE of the nlCWFS is approximately 0.3 waves for both data sets. 

Based on experience in the lab, we find that the two sensors fail in different ways. The SHWFS fails gradually as the number of focal plane spots that drop out increases with decreasing flux. For those remaining spots with sufficient signal, relatively accurate gradient measurements may still be made albeit only locally. Meanwhile, the nlCWFS fails more dramatically (abruptly) because the reconstruction algorithm depends on measuring interference fringes. When the four diffracted light signals become weak and less reliable, reconstruction solutions for the field rapidly lose self-consistency. This effect is then exacerbated by the phase unwrapping algorithm, which depends on receiving a reliable wavefront reconstruction. In the absence of an accurate phase estimate, the phase unwrapper frequently makes the reconstruction worse, resulting in WFE values that can exceed the original wavefront RMS.

\section{Summary and Conclusions}\label{sec:summary}
In this paper, we have studied the response of two different WFS's in mild and moderate scintillation. We find that the nlCWFS is intrinsically more sensitive than a comparable SHWFS, and also more robust to the effects of scintillation. In the next phase of the experiment, we will explore a broader range of scintillation indices to test the limits of the nlCWFS. Such experiments also need to be repeated in closed-loop operation.

\section*{ACKNOWLEDGMENTS}

This research was supported in part by the Air Force Office of Scientific Research (AFOSR) grant number FA9550-22-1-0435. JC acknowledges support from the Naval Research Lab (NRL) summer faculty fellowship program. Research presented in this article benefited from the use of Notre Dame's Center for Research Computing (CRC) and Engineering and Design Core Facility (EDCF). 

\bibliography{report}
\bibliographystyle{spiebib}

\end{document}